\documentstyle[prl,graphicx,twocolumn,aps]{revtex}
\begin{document}

\draft

\twocolumn[\hsize\textwidth\columnwidth\hsize\csname
@twocolumnfalse\endcsname

\title{Vortices in atomic-molecular Bose-Einstein
condensates}

\author{Tristram J. Alexander$^1$, Yuri S.  Kivshar$^1$, Elena A.
Ostrovskaya$^1$, and Paul S. Julienne$^2$}

\address{$^1$Nonlinear Physics Group, Research School of Physical
Sciences and Engineering, The Australian National University,
  Canberra ACT 0200, Australia \\
$^2$Atomic Physics Division, National Institute of Standards and
Technology, Gaithersburg, MD 20889}

\maketitle

\begin{abstract}
The structure and stability of vortices in hybrid atomic-molecular
Bose-Einstein condensates is analyzed in the framework of a
two-component Gross-Pitaevskii-type model that describes the
stimulated Raman-induced photoassociation process. New types of
topological vortex states are predicted to exist in the coherently
coupled two-component condensates even without a trap, and their
nontrivial dynamics in the presence of losses is demonstrated.
\end{abstract}

]

\narrowtext

\section{Introduction}

Recent experimental observation of the photoassociation of Bose
condensed atoms via the stimulated Raman process and the
production of ultracold molecules \cite{wynar} is, undoubtedly,
the first step towards the creation of a coherently coupled
atomic-molecular condensate. Such a system would represent a rich
ground for exploring a variety of the nonlinear atom-optics
effects, analogous to those known for the processes of parametric
up and down conversion of light waves \cite{up,down}. From this
point of view, the excited topological states of the macroscopic
wavefunction-- {\em vortices}-- in the atomic-molecular
Bose-Einstein condensates are of great interest as a direct
counterpart of the parametric vortices in nonlinear optics
\cite{Tristram,Isaac}. On the other hand, the study of such
topological states is important for understanding the properties
of composite superfluids, such as the recently proposed molecular
Bose-Einstein condensate (BEC) coupled to a cloud of Fermi atoms
\cite{bose-fermi}, or the composite atomic-molecular BEC (AMBEC)
\cite{ours}.

From the experimental point of view, it is important to  not only
establish the conditions for the formation of topological states,
but also identify the states which are dynamically stable, i.e.
are not destroyed by exponentially growing collective excitations
of the condensate. One of the proposed methods for the observation
of vortices involves imaging of the expanding condensate released
from a trap, with a vortex core growing during the expansion. The
atomic-molecular condensate represents an interesting experimental
challenge as the excited states may be formed in a dynamically
stable, self-confined droplet \cite{ours}, after the hybrid
condensate is released from a trap.

Assuming further experimental progress towards the creation of the
molecular and hybrid atomic-molecular BECs, in this paper we
analyse the dynamics of the parametrically coupled atomic and
molecular condensates, in the framework of a two-component
mean-field model \cite{model} that takes into account all types of
the mean-field atomic and molecular interactions \cite{ours}.  For
the first time to our knowledge, we study the structure,
stability, and dynamics of topological states in AMBEC, and, in
particular, we reveal that the dynamically stable {\em
two-component vortex solitons} can be formed in both trapped and
untrapped AMBEC clouds. Unlike the conventional vortices predicted
and already observed in atomic BEC, vortex solitons can exist as
stable {\em self-trapped states} even without a trapping
potential, supported by {\em an effectively attractive
interaction} between the parametrically coupled  atomic and
molecular components.

Another crucial problem is the dynamics of AMBEC and its
topological states in the presence of dissipation, i.e. loss of
atoms and molecules from the corresponding fractions of the hybrid
BEC. The formation of BEC in the presence of losses due to either
a coherent Raman photoassociation process or Feshbach resonance
has been analyzed previously \cite{Heinzen,Julienne}, however the
dissipative dynamics of vortices, earlier studied for a
single-component atomic BEC \cite{vortex_diss,ring_diss}, has
never been discussed in the context of hybrid condensates.

Since, so far, no conclusive experimental data are available that
would allow us to estimate the rates of the condensate losses
rigorously, in this work we take them into account by including {\em the
phenomenological dissipative terms}, and then considering the
dissipation rates for which the topological states can still be
formed. We analyze the dynamics of the vortex solitons in the
presence of dissipation and reveal that they can exhibit a
nontrivial decay scenario, with the decay rates much smaller than
the characteristic loss rates. Since the physical principles
underlying the existence and stability of these topological states
are rather generic, we expect that the similar structures can be
found in more realistic models of the atomic-molecular BEC
dynamics, and eventually observed in experiment.

\section{Model}

We study the dynamics of the hybrid atom-molecular condenstates,
produced by a coherent stimulated Raman photoassociation process
in a strongly anisotropic trap with a tight confinement along one
spatial dimension. We model the parametrically interacting
condensates by a system of two coupled Gross-Pitaevskii (GP)
equations for the macroscopic wave functions of the atomic
($\Psi_a$) and molecular ($\Psi_m$) components (see details and
references, e.g.,  in Ref. \cite{ours}):
\begin{eqnarray}
\label{model1}
    i\frac{\partial \Psi_a}{\partial t} + \frac{1}{2}\Delta\Psi_a
    -V_a\Psi_a -\chi \Psi_a^{*} \Psi_m e^{i\delta t} \nonumber \\
- (U_{aa}|\Psi_a|^2 + U_{am}|\Psi_m|^2) \Psi_a  = iR_a, \nonumber \\
i\frac{\partial \Psi_m}{\partial t} +
\frac{1}{4}\Delta\Psi_m - V_m\Psi_m - \frac{1}{2}\chi \Psi_a^2
e^{-i\delta t} \nonumber \\
 - ( U_{mm} |\Psi_m|^2 + U_{am}|\Psi_a|^2)\Psi_m = iR_m,
\end{eqnarray}
where the wave functions, time, and spatial coordinates are
measured in the units of $x_0^{-3/2}$, $\omega_{\perp}^{-1}$, and
$x_0 \equiv (\hbar/m\omega_{\perp})^{1/2}$, respectively.  Here
$\omega_{\perp}$ is the characteristic frequency in the plane
$(x,y)$ of the weak confinement of the trap for the atomic BEC
fraction. Consequently, $V_a = (1/2)r^2+(\lambda^2/2)z^2$ and $V_m
= r^2+\lambda^2z^2$, with $r^2=x^2+y^2$, and
$\lambda^2=\omega_z/\omega_\perp \gg 1$. In these notations, the
interaction strengths $U_{ij}$ are measured in the units of
$(\hbar\omega_{\perp})^{-1}x_0^{-3}$, and the Raman-induced
coupling, $\chi$, in the units of
$(\hbar\omega_{\perp})^{-1}x_0^{-3/2}$. The terms $R_{a,m}$ on the
right hand side are the phenomenological loss terms which will be
discussed and specified below in Sec. 4.2.

Due to significantly dissimilar scales of spatial confinement in the
different directions, and the {\em quasi-two-dimensonal}
geometry of the trap, the form of the condensate wavefunction in
the $z$ direction is completely determined by the structure of the
harmonic trap. We therefore assume that the wavefunction can be factorized in the
following way: $\Psi_{a,m}=\phi_{a,m}(x,y,t)\Phi_{a,m}(z)$, where
$\Phi_{a,m}(z)$ satisfy the one-dimentional harmonic oscillator equations:
\begin{equation}
\label{ho}
\frac{1}{2}
 \frac{\partial^{2} \Phi_{a,m}}{\partial z^{2}} -
 v_{a,m}(z) \Phi_{a,m}+\epsilon_{a,m}\Phi_{a} =0,
\end{equation}
where $v_{a}=(1/2)\lambda^{2}z^{2}$ and $v_{m}=2\lambda^{2}z^{2}$.
The ground state solutions of these equations correspond to
$\epsilon_{a}=\lambda/2$ and $\epsilon_{m}=\lambda$,
so that $\Phi_{a}=C_{a}\exp(-\lambda z/2)$ and $\Phi_{m}=C_{m}\exp(-\lambda
z)$. The normalization conditions $\int \Phi^{2}_{a,m}=1$ yield
 $C_{a}=(\lambda/\pi)^{1/4}$ and $C_{m}=(2\lambda/\pi)^{1/4}$.

Substituting the factorized wavefunctions into the model
equations, multiplying the first and second equations by
$\Phi_{a}$ and $\Phi_{m}$, respectively, integrating to eliminate
the $z$ dependence, and introducing
$\psi_{a,m}(x,y,t)=\phi_{a,m}(x,y,t)\exp(-i\lambda t/2)$, we
obtain the final evolution equations for the wavefunctions of a
``pancake''-shape AMBEC:
\begin{eqnarray}
\label{model}
    i\frac{\partial \psi_a}{\partial t} + \frac{1}{2}\Delta_{\perp}\psi_a
    -V_a(r)\psi_a -\tilde{\chi}\psi_a^{*} \psi_m e^{i\delta t} \nonumber\\
- \left( \tilde{U}_{aa}|\psi_a|^2 +
\tilde{U}_{am}|\psi_m|^2 \right) \psi_a  = i\tilde{R}_a, \nonumber \\
    i\frac{\partial \psi_m}{\partial t} +
\frac{1}{4}\Delta_{\perp}\psi_m - V_m(r)\psi_m - \frac{1}{2}\tilde{\chi}\psi_a^2
e^{-i\delta t} \nonumber \\
 - ( \tilde{U}_{mm} |\psi_m|^2 +
 \tilde{U}_{am}|\psi_a|^2)\psi_m = i\tilde{R}_m,
\end{eqnarray}
where $\Delta_{\perp}=\partial^{2}/\partial
x^{2}+\partial^{2}/\partial y^{2}$,  $V_{a}(r)= (1/2)r^2$, and
$V_{m}(r)= r^2$.  Here we introduce the renormalized interaction
strengths: $\tilde{U}_{aa}=U_{aa}(\lambda/2\pi)^{1/2}$,
$\tilde{U}_{mm}=U_{mm}(\lambda/\pi)^{1/2}$,
$\tilde{U}_{am}=U_{am}(2\lambda/3\pi)^{1/2}$,
$\tilde{\chi}=\chi(\lambda/2\pi)^{1/4}$ and, correspondingly, the
renormalized loss terms $\tilde{R}_{a,m}$. For the sake of
clarity, we omit the tilde from the equations throughout
 the following text. One can see that the renormalization factors
 absorbed by the interaction coefficients as a result of the reduction
 to the 2D geometry are, in fact, close to $1$ for typical $\lambda
 \sim 10$.

 For definiteness, we consider the case of $^{87} Rb$, for which
the production of ultracold molecules were recently demonstrated
experimentally \cite{wynar}, and assume that the Raman detuning
parameter vanishes, $\delta=0$. Using the recently reported data
for the intra- and inter-species scattering lengths \cite{wynar},
we take $U_{aa} = 0.062$; $U_{am} = -0.084(\pm 0.07)$, and $\chi =
1.09$. The large uncertainty in the value of $U_{am}$ corresponds
to the s-wave scattering length, $a_{am} = -180 a_0 \pm 150 a_0$,
recently measured experimentally \cite{wynar}. Since the molecular
BEC has not been experimentally realized yet, the exact value of
$U_{mm}$ is not known, although it can be estimated to be of the
order of the other interaction strengths (see details in Ref.
\cite{ours}) and, therefore,  it cannot be ignored in the model
(\ref{model}).

It has also been previously established that the qualitative
behaviour of the system depends on the sign of the determinant
$\Delta U\equiv U_{aa}U_{mm}-U^2_{am}$ \cite{ours,model}, which
defines the relative strength of the cubic interactions. Parameter
$U_{mm}$ can therefore be used to investigate the system in two
distinct regimes of the net {\em attractive} $\Delta U>0$ and net
{\em repulsive} $\Delta U<0$ cubic interactions \cite{ours}. At
the low particle numbers, i.e. for the low densities of the
condensate wavefunctions, and in the parameter region where
$\Delta U \approx 0$, the interactions are dominated by the
parametric, i.e. quadratic, coupling ($\sim \chi$) which can be
{\em effectively attractive or repulsive}, depending on the
relative phase of the molecular and atomic components \cite{ours}.
The structure and dynamical properties of all topological
structures in the AMBEC is controlled by the complex interplay of
the quadratic and cubic inter- and intra-component mean-field
interactions.

\section{Vortex solitons}

Spatially localized ground-state solutions of the model
(\ref{model}) at $R_a=R_m=0$, with and without the trapping
potential $V_j(r)$,  have been analyzed earlier \cite{ours}. Here,
we are interested in the excited states of the model (\ref{model})
that correspond to the topological vortex states of the hybrid
two-component condensate. In order to find the vortex states,
we consider the model equations (\ref{model}) {\em without
the dissipative terms} (i.e. $R_a=R_m=0$), and look for the
radially symmetric solutions with a nontrivial phase in the form:
\begin{eqnarray}
\label{stat} \psi_{a}(r, \theta,t) =\phi_{a}(r)
e^{ik_{a}\theta-i\mu_{a}t},\nonumber \\
\;\;\; \psi_{m}(r, \theta,t)=\phi_{m}(r)
e^{ik_{m}\theta-i\mu_{m}t},
\end{eqnarray}
where $(r,\theta)$ are the polar coordinates in the plane of the
trap, and $\mu_{a,m}$ are the chemical potentials of the
corresponding AMBEC fractions. Due to the parametric coupling between the atomic and
molecular matter waves, the existence of {\em stationary states}
of the AMBEC requires the phase-matching conditions,
$\mu_{m}=2\mu_{a}\equiv 2\mu$, and $k_{m}=2k_{a}
\equiv 2n $, so that the integer $n$ defines the topological charge of
the AMBEC vortex soliton via its atomic component.

\begin{figure}
  \centerline{\includegraphics[width=8.0cm]{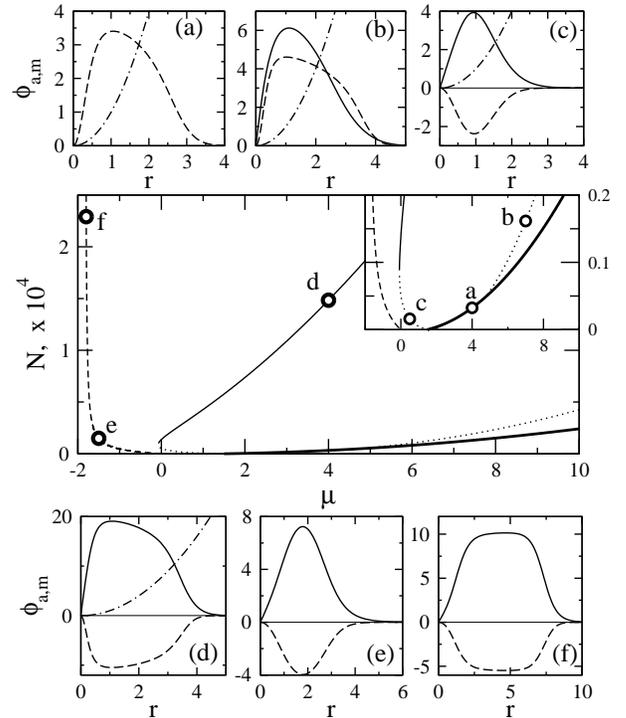}}
  \caption{Middle: Different families of the stationary one- and two-component vortex states
  in the model (\ref{model}).
Thin solid/dashed - stable/unstable hybrid vortices in a trap,
dotted - unstable vortices in a trap, dashed - untrapped
self-localized vortex solitons. Thick solid - vortices of a
molecular condensate. Top and bottom: Examples of the
stationary profiles of the vortices corresponding to the marked
points. Dashed-dotted curves in (a)-(d) show the trapping
potential. Insert: Close-up of the bifurcation region.}
  \label{fig2}
\end{figure}

The spatially localized, radially symmetric solutions of the model
(\ref{model}) in the form (\ref{stat}) can be found by a standard
numerical relaxation technique. Different types of such
topological states are summarized in Fig. \ref{fig2} for the case
$n=1$ (single-charged vortices). The upper solid curve in the
middle part of  Fig. \ref{fig2} [that includes the point (d)]
corresponds to {\em stable vortices in a trapping potential}; such
vortices are supported by the effectively {\em repulsive cubic
interaction} since $\Delta U < 0$. These topological states are
somewhat similar to the vortices of a single-component atomic BEC,
and they expand when released from the trap. At the low particle
numbers [see the inset in Fig. \ref{fig2}, dashed curve with the
point (c)], these vortex solutions become unstable due to the
competition between the effectively {\em attractive quadratic} and
{\em repulsive cubic} interaction. Thus, the repulsive nature of
the inter-species cubic interaction manifests itself only for
large densities of the components $\phi_a$ and $\phi_m$.

Thick solid curve in Fig. \ref{fig2} (and the insert) describes a
family of the vortices of the single-component {\em purely
molecular} condensate wavefunction $\psi_m$ [see the case (a)],
whereas a dotted curve above it corresponds to another family of
{\em hybrid vortices} that bifurcates from the purely molecular
states. This second family of the two-component vortices
is supported by an effectively {\em repulsive quadratic}
nonlinearity allowing such states to exist for low particle
numbers but only in the presence of a trapping potential [cf. an
example in (b)].

An important feature of the model (\ref{model}) is the existence
of untrapped self-localized {\em vortex solitons} for $\mu < 0$.
These structures are supported by an effectively {\em attractive
quadratic} interaction between the atomic and molecular
components. As the parameter $\mu$ decreases, the solution becomes
broader due to the presence of effectively repulsive cubic nonlinearity, with
a saturation amplitude of the wavefunction corresponding to the amplitude
of the spatially homogeneous (free-space) condensate wavefunction
[see the cases (e) and (f) in Fig. \ref{fig2}].

In this paper, we determine the vortex stability by a direct
numerical propagation, but a more rigorous stability analysis will
be presented elsewhere. We find that the broad vortex solitons
[shown, e.g., in Fig. \ref{fig2}(f)] are {\em stable}, whereas
narrow vortex solitons [e.g. those shown in Fig. \ref{fig2}(e)]
are {\em unstable}, in agreement with the earlier analysis of the
parametric vortices in optics \cite{Isaac}. Interestingly, for the
unstable vortices we observe two types of the
instability-induced dynamics, i.e. the break-up into {\em
two} or {\em three} filaments, for larger and smaller particle
numbers, respectively.  This dynamical behaviour is a clear signature
of the transition between the
regime of effectively cubic interaction in which the vortex
soliton decays into $2|n|$ filaments \cite{firth}, and the regime
of effectively quadratic interaction when the vortex soliton
breaks up into $2|n| +1$ filaments, as usually observed for
quadratic parametric beams \cite{chi2}.

\begin{figure}
  \centerline{\includegraphics[width=8.4cm]{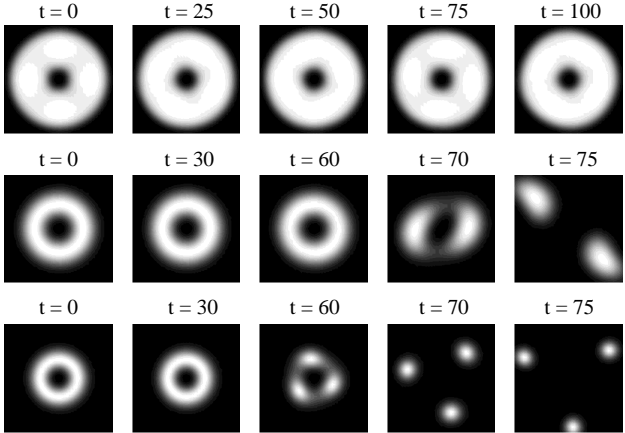}}
  \caption{Lossless dynamics of the self-trapped vortex
  soliton. Top: stable propagation of a broad vortex, the case (f)
  in Fig. \ref{fig2} ($\mu =-1.8$). Middle: the symmetry-breaking instability
  induced by the effective cubic nonlinearity, $\mu =-1.5$, the case (e) in Fig. 1.
  Bottom: the symmetry-breaking instability induced by parametric
  interaction, $\mu = -0.5$. Only the atomic component is shown. }
  \label{fig2a}
\end{figure}

\section{Dissipation-induced vortex dynamics}

\subsection{Physical estimates of the losses}

In the formation of ultracold molecules from an atomic BEC, the
processes of the incoherent production of molecules, decays of the
atomic excited state, and inelastic collisions between atoms and
molecules, all contribute to the the effective loss of particles
from the corresponding fractions of the BEC, resulting in the
dissipative dynamics of the localized AMBEC states.

To take into account all possible losses in the hybrid condensate,
we introduce the following linear and nonlinear (i.e.,
density-dependent) dissipative terms into the GP model
(\ref{model}),
\begin{equation}
\label{losses}
\begin{array}{l} {\displaystyle
 R_a= - \nu_{a}\psi_{a} -
\gamma_{a}|\psi_{a}|^{2}\psi_{a} -
\gamma_{am}|\psi_{m}|^{2}\psi_{a}, }\\*[9pt] {\displaystyle R_m= -
\nu_{m}\psi_{m} - \gamma_{m}|\psi_{m}|^{2}\psi_{m} -
\gamma_{am}|\psi_{a}|^{2}\psi_{m}, }
\end{array}
\end{equation}
where the dimensionless coefficients, $\nu_{j}$ and $\gamma_{j}$, stand
for the decay rates of the atomic excited state ($\sim \nu_{a}$)
and the excited molecules into high bound states of the atomic BEC
or ``hot'' (non-BEC) atoms ($\sim \nu_{m}$), the rates of
inelastic collisions and decay of the atomic
states ($\sim \gamma_{a}$), and inelastic molecule-molecule
collisions and dissociation into high bound state or continuum
state of the molecular BEC ($\sim \gamma_{m}$), and atom-molecule
inelastic collisions resulting in the production of ``hot'' atoms
and molecules ($\sim \gamma_{am}$).

The linear loss terms in Eq. (\ref{losses}) are similar to those
discussed earlier in Ref.  \cite{Heinzen}, however the latter work
did not take into account the losses due to the rotationally or
vibrationally inelastic atom-molecule collisions, as well as the
molecule-molecule collision losses. System (\ref{model}) with the
loss terms (\ref{losses}) is formally equivalent to the
mathematical description of the formation of AMBEC through
Feshbach resonances \cite{Julienne}, except for the terms
responsible for the linear losses from the excited atomic state
($\sim \nu_{i}$) which are specific for the AMBEC production via
the coherent Raman photoassociation process. We stress that no
conclusive experimental data for the losses that would enable the
coherent AMBEC production are available at the moment, so some
reasonable estimates need to be made.

In Eqs. (\ref{losses}), the coefficients for the collisional
losses, $\gamma_j$, are measured the units of
$\Gamma=x^3_0\omega_\perp$, and for the typical values of $x_0\sim
5$ $\mu \rm{m}$, and $\omega_\perp\sim 10^2$ ${\rm Hz}$,
$\Gamma\sim 10^{-8}$ ${\rm cm^3/s}$. In what follows,  we
investigate the effect of relatively ``weak'', $\gamma_j=10^{-3}$
and ``strong'', $\gamma_j=10^{-1}$ losses, which nevertheless
corresponds to physically ``strong'' inelastic collisions rates of
$\Gamma=10^{-11}-10^{-9}$ ${\rm cm^3/s}$. For simplicity, we
assume that all the inelastic loss rates in the model are equal to
each other.

As was estimated in Ref. \cite{Heinzen}, the linear loss rates
that still enable the coherent dynamics of the AMBEC are of the
order of $ 10^2$ ${\rm s}^{-1}$, which corresponds to our
dimensionless values of $\nu_j\sim 1$. Assuming this to be the
case, we note that the term responsible for the ``weakest''
nonlinear losses in our model still an order of magnitude larger
than the corresponding linear loss, i.e. $\Gamma n_j \sim 10^3$
${\rm s}^{-1}$ for the typical condensate density of $n_j\sim
10^{14}$ ${\rm cm}^{-3}$.  This means that, without loss of
generality, we can assume that $\gamma_j=\nu_j$ in our model, and
bear in mind that the effect of the linear losses is only
significant when the particle numbers (and condensate densities)
are small.

The other (strong) loss mechanism for the molecular fraction,
which is absent in our effectively two-mode consideration, is the
coupling of a molecule to the other, excited states of the atomic
BEC due to the closeness of the excited molecular level to the BEC
atom-pair level. This loss channel has been recently shown to be
responsible for the enhancement of the condensate loss rates
\cite{yurovsky}. One can argue, however, that on the time scales
large compared with the corresponding rates of decay into the
higher-order modes, the lowest-order nonlinear excited modes of
the hybrid AMBEC that we consider here become decoupled from the
rest of the bound state spectrum, and all losses associated with
such coupling can be effectively incorporated into the
phenomenological terms proportional to $~\nu_m$.

\subsection{Lossy dynamics of the vortex soliton}

In the presence of losses, the total particle number in both
components of AMBEC decreases. First of all, when the molecular
component is small, the effective equations (\ref{model}) can be
decoupled and the dissipative dynamics of two components can be
described separately. In particular, neglecting the molecular
component, and averaging over the spatial structure of the atomic
BEC wavefunction, from Eqs. (\ref{model}),(\ref{losses}) we obtain
the simple equation for the density $n_a =|\psi_a|^2$ of the
atomic component
\begin{equation}
\label{decay} i\frac{dn_a}{dt} = - 2 \nu_a n_a - 2 \gamma_an_a^2,
\end{equation}
which has the analytic solution,
\begin{equation}
\label{anal}
 n_a(t) =\frac{\nu_a e^{-2\nu_a t}}{( C -\gamma_a
e^{-2\nu_a t})},
\end{equation}
where $C = \gamma_a + \nu_a /n_a(0)$. When the effect of linear
losses can be neglected, the decay of the atomic component follows
a power law.

\begin{figure}
\centerline{\includegraphics[width=7.5cm]{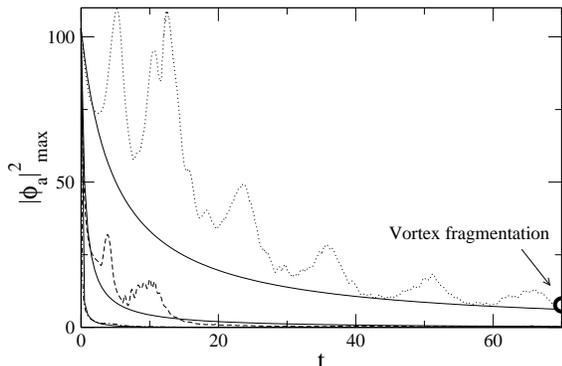}}
\caption{Decay of the atomic component
amplitude for different values of losses. The initial state is a
self-trapped vortex soliton as in Fig. 1(f) and Fig. 2(top).
Losses are: 0.001 (dotted), 0.01 (dashed), and 0.1 (dash-dotted).
Thin solid lines show the analytic dependence (\ref{anal}).}
\label{figNEW}
\end{figure}

We have conducted the numerical simulations of the
dissipation-induced dynamics of the vortex soliton varying the
loss terms. In Fig. \ref{figNEW}, we show the decay of the maximum
amplitude of the atomic component $n^{\rm max}_a=|\psi_a|^2_{\rm
max}$ of the two-component vortex soliton in the presence of the
losses of a different strength. For ``strong'' and ``medim''
losses ($\gamma=\nu=0.1$ and $0.01$ respectively), the decay
follows approximately the analytical dependence (\ref{anal}) and,
therefore,  the modal (spatial) structure of the vortex state is
not very important.

\begin{figure}
  \centerline{\includegraphics[width=7.5cm]{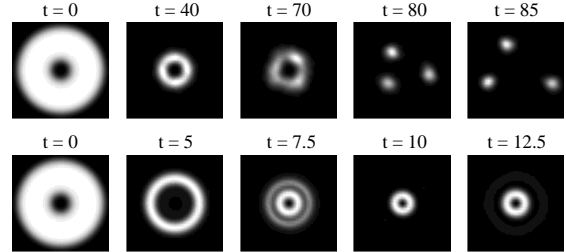}}
  \caption{Dynamics of the (initially) stable vortex soliton in the presence
  of  losses: (a) weak, $\gamma_{i}=\nu_{i}=0.001$, and (b) strong, $\gamma_{i}=\nu_{i}=0.01$.}
  \label{fig4}
\end{figure}

For weaker losses, we reveal novel features in the vortex dynamics
associated with the interplay between the instability-induced
vortex fragmentation and the decay of the vortex amplitude due to
losses.  As an example, in Fig. \ref{fig4}(top) we show the
propagation of the initially stable, untrapped, vortex soliton
[shown as (f) in Fig. 1] in the presence of weak dissipation,
$\gamma_i = \nu_i = 0.001$. While this vortex is absolutely stable
in the lossless regime (see Fig. 2, top), its number of particles
decreases and the corresponding vortex solution moves along the
dashed curve in Fig. 1 from the point (f) to the point (e),
entering the lower part of the branch corresponding to unstable
vortices. Then, the vortex fragmentation into two or three
filaments occurs, as shown in Fig. \ref{fig4}(top). This leads us
to conclude that the fragmentation to components occurs when the
particle number has dropped far enough to place the vortex in the
region of instability for stationary solutions and, overall, the
vortex parameters adiabatically follow the family of untrapped
localized solutions.

For comparison, in Fig. \ref{fig4}(bottom) we show the vortex
dynamics for stronger losses, when no fragmentation is observed at
all. In this case, the vortex soliton does not follow a family of
localized solutions adiabatically. Moreover, it appears that the
dissipation leads to a switching from the fundamental vortex state
to one of the higher-order vortex modes.  The additional ``rings''
in the density distribution observed in Fig. \ref{fig4} (bottom)
therefore seem to be a manifestation of such a higher-order mode.

\subsection{Evolution of a vortex released from a trap}

Although different possibilities to create the topological states
in experiment may exist, it is probably safe to assume that the
presence of the confining trap is paramount. The question
therefore is whether a dynamically stable topological state of
AMBEC created in a trap can be transformed into a dynamically
stable vortex in a self-confined AMBEC droplet, after the hybrid
condensate is released from the trap. We investigate such a
possiblity by conducting additional numerical studies.

\begin{figure}
  \centerline{\includegraphics[width=7.5cm]{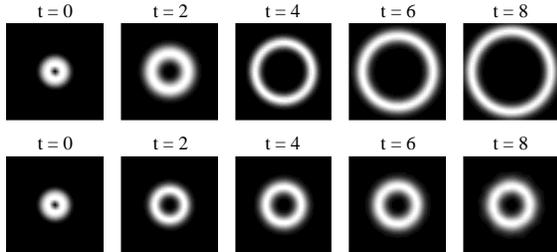}}
  \caption{Dynamics of a vortex state formed in a
  trap and released from the trap at $t= 0$, under the influence of
  (a) weak ($\gamma=\nu=0.001$) and (b) strong ($\gamma=\nu=0.01$) losses.}
  \label{fig6}
\end{figure}

 A trapped vortex, corresponding to the stable family shown in the
main part of  Fig. 1 [the branch that includes the point (d)] is
released from the trap at $t =0$. We find that for weak or no
dissipation this vortex structure diffracts rapidly, as shown in
Fig. \ref{fig6}(top). However, when the losses become large
enough, the released vortex state gets {\em self-trapped}, as
shown in Fig. \ref{fig6}(bottom).

\begin{figure}
  \centerline{\includegraphics[width=7.5cm]{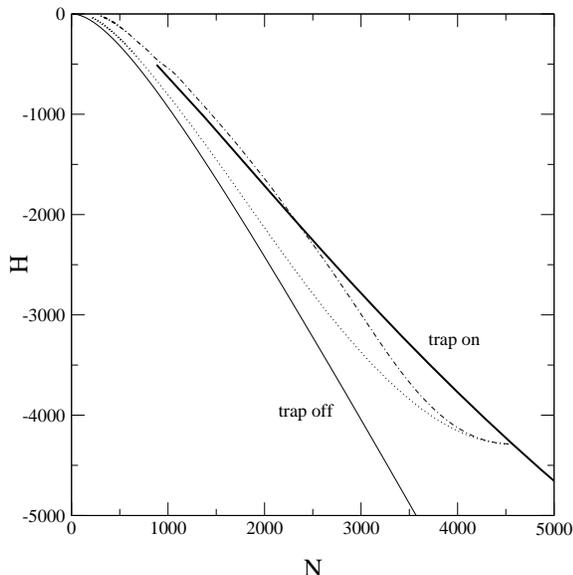}}	
  \caption{Parametric plot $H(t)$ vs. $N(t)$ for the vortex
  evolution shown in Fig. \ref{fig6}. Solid (thick and thin) lines - the families
  of the vortex states with and without the trap, respectively. Dashed and dotted curves show
  the dynamics of the stable vortex initially formed in a trap and released from it
  in the presence of losses,
   $\gamma=\nu=0.001$ and $\gamma=\nu=0.01$,  respectively.}
  \label{fig7}
\end{figure}

In order to verify the effect of self-trapping and get a deeper
insight into our results, we use the parametric plot of the system
Hamiltonian $H(t)$ vs. the total particle number $N(t)$, which both
evolve in time in the presence of losses. For weak losses,  we
reveal that the vortex dynamics follows the ``trapped'' family of
the vortex states (dashed line in Fig. \ref{fig7}). The family of
the trapped vortex states in this plot corresponds to the initial
conditions of a trapped vortex stationary solution with the
trapping potential turned off.  This vortex state diffracts when
it is released from the trap due to the dominant repulsive cubic
nonlinearity. However, for stronger losses we observe a novel
effect when the vortex,  being initially  released from the trap,
does not rapidly diffract but, instead, it switches to a
self-trapped vortex soliton, as shown by a dotted curve in Fig.
\ref{fig7}.

\section{Conclusions}

In the framework of a two-component mean-field model of hybrid
coherently coupled condensates, we have analyzed the structure,
stability, and dynamics of the topological states in the
atomic-molecular BEC created via the induced two-color Raman
photoassociation process. We have predicted the existence of novel
two-component vortex solitons supported by the coherent parametric
coupling between the atomic and molecular BEC fractions, and
demonstrated that such topological states can exist even in the
absence of a trapping potential.

Since the losses are known to be very important in the formation
of the hybrid atomic-molecular condensates, we have analyzed the
effect of weak and strong losses of a different physical origin on
the dynamics of the vortex solitons, the type of problems never
addressed in other fields such as nonlinear optics. In particular,
for the first time to our knowledge, we have described an
interplay between the nonlinearity-induced vortex fragmentation
and dissipation-induced vortex decay. Moreover, we have revealed
that the losses may be very useful to identify the self-trapped
vortex states in the hybrid condensates in experiment, after
releasing the atomic-molecular condensate from a trap.

Although our analysis is valid for a quasi-two-dimensional model,
the generalization to a spherically symmetric case seems
straightforward. Additionally, in spite of the limited validity of
the two-component model, we believe that the similar types of
topological states can be found in more realistic models of hybrid
condensates, as well as in other types of coherently coupled
mixtures of the multi-component condensates such as spinor
Bose-Einstein condensates.

The work has been supported by the Planning and Performance
Fund of the Institute of Advanced Studies at the Australian
National University.

\end{document}